\documentclass{article}
\usepackage{ijcai22}

\usepackage{times}
\usepackage{soul}
\usepackage{url}
\usepackage[hidelinks]{hyperref}
\usepackage[utf8]{inputenc}
\usepackage[small]{caption}
\usepackage{graphicx}
\usepackage{amsmath}
\usepackage{amsthm}
\usepackage{booktabs}
\usepackage{algorithm}
\usepackage{algorithmic}
\title{Agents Incorporating Identity and Dynamic Teams in Social Dilemmas}

\author{
Kyle Tilbury
\and
Jesse Hoey
\affiliations
David R. Cheriton School of Computer Science, University of Waterloo
\emails
\{ktilbury, jesse.hoey\}@uwaterloo.ca
}

\begin{document}

\maketitle

\begin{abstract} 
We present our preliminary work on a multi-agent system involving the complex human phenomena of identity and dynamic teams.  
We outline our ongoing experimentation into understanding how these factors can eliminate some of the naive assumptions of current multi-agent approaches. 
These include a lack of complex heterogeneity between agents and unchanging team structures. We outline the human social psychological basis for identity, one’s sense of self, and dynamic teams, the changing nature of human teams. 
We describe our application of these factors to a multi-agent system and our expectations for how they might improve the system’s applicability to more complex problems, with specific relevance to ad hoc teamwork. 
We expect that the inclusion of more complex human processes, like identity and dynamic teams, will help with the eventual goal of having effective human-agent teams.
\end{abstract}

\section{Introduction}


Multi-agent systems, such as in multi-agent reinforcement learning (MARL) and ad hoc teamwork (AHT), have enabled us to study teamwork between agents \cite{durugkar2020balancing,genter2011role,mirsky2022survey,hughes2018inequity,leibo2021scalable,baker2019emergent,jaderberg2019human,leibo2017multi}.
However, due to multi-agent settings being inherently difficult, past work tends to make various assumptions lowering the applicability to the real world.
We extend multi-agent settings to include more complex real-world aspects, specifically social identity and dynamic teams. 

In human groups there is a complex relationship between how group (team) membership shapes social identity and how social identity shapes team membership~\cite{schroder2016modeling}. We need to consider more complex human processes in agent systems if we want to have effective human-agent teams.
Humans have diverse identities, their sense of self and of one's position in a social and cultural structures~\cite{fukuyama2018identity}. 
Current multi-agent approaches tend to regard agents as homogeneous or slightly heterogeneous. Similarly to humans, agents should be able to contend with complex social phenomena like identity to perform well in the real world.
Current multi-agent reinforcement learning team approaches are too naive; agents are typically assigned an explicit team. 
Human teams are fluid and changing and the agents teams need to encompass these attributes in their own teams.
We suggest more human like team assignment, where agents can have the ability to choose their teams and personal preferences towards tasks. 

We illustrate how the inclusion of these complex aspects can be investigated in sequential social dilemmas~\cite{leibo2017multi}. 
These are multi-agent games that are partially observable, and temporally and spatially extended. They are social dilemmas in the sense that an individual agent is motivated to exploit non-cooperative strategies for short-term gain, but in the long run all agents would have a higher payoff if they all cooperated.

In our work, we utilise the popular Markov game known as Cleanup~\cite{hughes2018inequity}, though there are many pre-existing multi-agent environments~\cite{leibo2021scalable}.
Cleanup is a sequential social dilemma that consists of a number of agents wherein agents are rewarded for picking apples.
The apples grow in an orchard at a rate that is inversely proportional to the amount of pollution in a nearby river. 
When the level of pollution in the river is over a certain threshold, apples cease to grow at all.
Agents' actions consist of moving around the environment, picking apples in the orchard, or cleaning pollution in the river. 
Agents must leave the orchard to clean the river. 
Success in Cleanup depends on the agents' ability to form a joint policy to effectively clean the river so that apples grow and they share in their consumption. In other words, agents have to coordinate picking apples with the behaviours of other agents in order to achieve cooperation. The collective reward obtained by a group of agents gives insights into how well the group learned to cooperate.

We now outline the basis of two complex human phenomena and detail our approach into applying them in Cleanup. This is followed by a short discussion on the implications of our work.

\section{Identity}

A person's identity affects economic outcomes. We consider identity economics~\cite{akerlof2000economics,akerlof2010identity}, an economic model of human behaviour that incorporates the psychology and sociology of identity. 
Identity economics captures classic economic utility, such as a person's tastes for goods and services, and also identity elements that capture how motivations vary with social context. These identity elements are: the \emph{social categories} and each individuals category assignment (\emph{identity}), the \emph{norms} and \emph{ideals} for each category, and the \emph{identity utility} corresponding to the gain when actions conform to norms/ideals and loss when they do not.

Multi-agent reinforcement learning work deals with predominately economic factors stemming from how the learning environment and reward functions are structured. So, we endow agents with identity components to accompany the economic components already present in Cleanup, drawing from identity economics. 

It is established that heterogeneous agents with their own individual preferences, for example when agents are partially selfish, can have better performance than purely selfless task-oriented agents~\cite{durugkar2020balancing}. 
Similarly with mixed objectives in AHT all team members can have a common goal, but each agent might also hold individual goals~\cite{mirsky2022survey}. 
However, we build upon these notions connecting to the social psychology concept of identity to further increase the heterogeneity of agents.

Within Cleanup, we employ agents with identities. We model identities as innate, but unknown, specializations that agents are randomly assigned and need to discover themselves over the course of the game, distinct from past work involving roles~\cite{wang2020roma}.
We are experimenting with specific identity specializations (social categories) with distinct identity-aligned incentives (identity utility) for agents. 
For agents whose identity is \emph{river cleaner} their expertise is in performing river cleaning actions. They clean more pollution from the river at once than does an agent whose identity is not river cleaner. 
For agents whose identity is \emph{apple picker} their expertise is in the action of harvesting apples. They can harvest more apples from the orchard at one time than does an agent whose identity is not apple picker.
If agents can discover and learn the norms and ideals to behave inline with their identities, in conjunction with forming teams as we discuss next, we expect them to outperform agents that do not incorporate identity.

\section{Dynamic Teams}

Human teams are dynamic. People join and switch teams with regard to their identity, common motives and goals, accepted division of labour or roles, status relationships, and accepted norms within the group~\cite{sherif2015group}. 
Past work has shown MARL teams can perform well and learn complex behaviours~\cite{baker2019emergent,jaderberg2019human}. 
In Cleanup specifically, teams have been shown to perform better overall than independent goal-aligned agents acting cooperatively~\cite{radke2022exploring,radke2022importance}. 
However, in these previous works, teams are static and lack the dynamic nature of human teams. With this in mind, we investigate what we call dynamic teams. 

With dynamic teams agents are capable of creating and switching teams or belonging to no team at all. 
If agents form a team, they must be able to coordinate and collaborate with their new teammates.
This can be viewed as a sort of fluid ad hoc teamwork where agents can enter or leave teams, but stay in the environment.
This contrasts with open ad hoc teamwork where agents may enter or leave the environment, but not individual teams~\cite{rahman2021towards}.
We want to understand under what conditions agents, balance their individual identities and group membership, come to form, switch, and stay in teams. 

Extending Cleanup, we give agents the action to choose a team.
We experiment with different allowable intervals of team switching and with locking agents into a team after a certain amount of time or amount of team changes has passed.
Agents can be in a team of size 1 to $n$, where $n$ is the total number of agents. So, agents must learn to account for this dynamic size of teams.
This is similar to the open environment problem in AHT where the learner has to adapt to the changing number of teammates~\cite{mirsky2022survey}. 
When agents are in a team together, they share all reward equally among each member of the team.
This reward sharing can allow agents to specialize based on their identity.
For example, river cleaner agents are free to spend the entirety of their time cleaning the river, whereas when they are alone, they must pick apples otherwise they receive no reward.
We expect that this specialization will lead to more optimal outcomes as was shown in role based ad hoc teamwork~\cite{genter2011role}.
Investigating dynamic teams could allow MARL methods to overcome weaknesses highlighted in past work where good team performance comes at the expense of poor performance when encountering a new teammate~\cite{rahman2021towards}.

\section{Discussion}
We envision interesting dynamics arising from the incorporation of identity and dynamic teams in Cleanup.
The addition of identity by itself, as we've structured it, will make the social dilemma more difficult to overcome as agents with apple picker identities are further incentivized to engage in non-cooperative behaviour. 
We are interested in the dynamics that arise due to different distributions of agent identities within the population.
We imagine that the addition of identity and teams will, if the agents can learn to take advantage of them, have better outcomes for the population as a whole. 
It is clear that a team comprised of agents with apple picker identities picking apples and agents with river cleaner identities cleaning the river will perform better collectively. 
However, it is not clear if it will be possible for agents to actually learn this.
Furthermore, we are interested in what size of teams emerge and the ratio of agent identities within the teams.
It may be desirable to train agents with one model, but multiple policies depending on the properties of the team that they are in.
For example, an agent could employ a different policy being in a team of size four than the policy the policy they would use when alone. 
Exploring these complex dynamics is an necessary precursor for reducing the difference between these toy environments and the real world.
With the ultimate goal of having effective human-agent teams, we need agents that can contend with more complex human phenomena rooted in social psychology.

\bibliographystyle{named}
\bibliography{ijcai22}

\end{document}